\documentclass[a4paper]{spie}

\usepackage[sc, hang]{caption}
\usepackage{graphicx}
\usepackage{graphics}
\usepackage{makeidx}
\usepackage[pass]{geometry}
\usepackage{booktabs}
\usepackage{marvosym}
\usepackage[hang]{subfigure}
\usepackage{SIunits}
\usepackage{rotating}
\usepackage{csquotes}
\usepackage{url}
\usepackage{multicol}
\usepackage{multirow}
\usepackage{overpic}
\usepackage{amsmath}
\usepackage{commath}
\usepackage{bm}
\usepackage{rotating}
\usepackage{psfrag}
\usepackage{parallel}
\usepackage{fancyhdr}
\usepackage[breaklinks]{hyperref}

\usepackage[english]{babel} 
\usepackage{eso-pic,graphicx}
\usepackage{enumerate}
\usepackage{psfrag}
\usepackage{booktabs}
\usepackage{hyperref}      
\hypersetup{               
   pdftitle={maorySPIE2020},    
   pdfauthor={Guido Agapito},  
   colorlinks=true,      
   linkcolor=black,      
   anchorcolor=black,    
   citecolor=black,      
   filecolor=black,      
   menucolor=black,      
   pagecolor=black,      
   urlcolor=blue,         
   linktocpage=true
}

\title{MAORY AO performances} 

\author[a,c]{Guido Agapito}
\author[a,c]{Cedric Plantet}
\author[a,c]{Lorenzo Busoni}
\author[b,c]{Carmelo Arcidiacono}
\author[d]{Sylvain Oberti}
\author[d]{Christophe Verinaud}
\author[d]{Miska Le Louarn}
\author[a,c]{Alfio Puglisi}
\author[a,c]{Simone Esposito}
\author[e]{Paolo Ciliegi}
\affil[a]{INAF Osservatorio Astrofisico di Arcetri, L. Enrico Fermi 5, 50125 Firenze, Italy}
\affil[b]{INAF Osservatorio Astronomico di Padova, Vicolo dell’Osservatorio 5, 35122, Padova, Italy}
\affil[c]{ADaptive Optics National laboratory in Italy (ADONI)}
\affil[d]{European Southern Observatory (ESO),Karl-Schwarzschild-Str. 2, D-85748 Garching bei Muenchen, Germany}
\affil[e]{INAF Osservatorio di Astrofisica e Scienza dello Spazio di Bologna, via Gobetti 93/3, Bologna, Italy}

\authorinfo{Further author information:\\Guido Agapito, \Letter
 \hspace{0.5ex} guido.agapito@inaf.it\\}

\begin{document} 

  \maketitle

  \begin{abstract}
    The Multi-conjugate Adaptive Optics RelaY (MAORY) should provide 30\% SR in K band (50\% goal) on half of the sky at the South Galactic Pole. Assessing its performance and the sensitivity to parameter variations during the design phase is a fundamental step for the engineering of such a complex system. This step, centered on numerical simulations, is the connection between the performance requirements and the Adaptive Optics system configuration. In this work we present MAORY configuration and performance and we justify the Adaptive Optics system design choices.
  \end{abstract}


\section{INTRODUCTION}\label{sec:intro}

MAORY (Multi-conjugate Adaptive Optics RelaY) for the Extremely Large Telescope (ELT)\cite{10.1117/12.2234585,paolo2020maory,mess2020} will be one of the first Multi Conjugate Adaptive optics (MCAO) system for the next generation of telescope.
Its design is a challenge and numerical simulations have been used as a fundamental support to the engineering of such a system.
Assessing the Adaptive Optics (AO) performance and the sensitivity to parameters variations is not trivial because the main metric is the statistical sky coverage.
Hence, a huge amount of work have been devoted to develop the sky coverage estimation library described in Ref.~\citeonline{10.1117/12.2313175} and \citeonline{Plantet2019} and to improve the end-to-end AO system simulator PASSATA\cite{doi:10.1117/12.2233963}.
The changes made to PASSATA are briefly described in Appendix \ref{sec:appendix}.
Thanks to these tools we estimate the baseline performance of MAORY in several atmospheric conditions (see Sec.\ref{sec:base}) and the sensitivity to the variation of some parameters (see Sec.\ref{sec:sens}).



\section{BASELINE PERFORMANCE}\label{sec:base}

\begin{figure}[h!]
\centering
\subfigure[K band SR as a function of the off-axis angle for different atmospheric conditions, for a single post focal DM and two post focal DMs.\label{fig:SR}]
{\includegraphics[width=0.5\columnwidth]{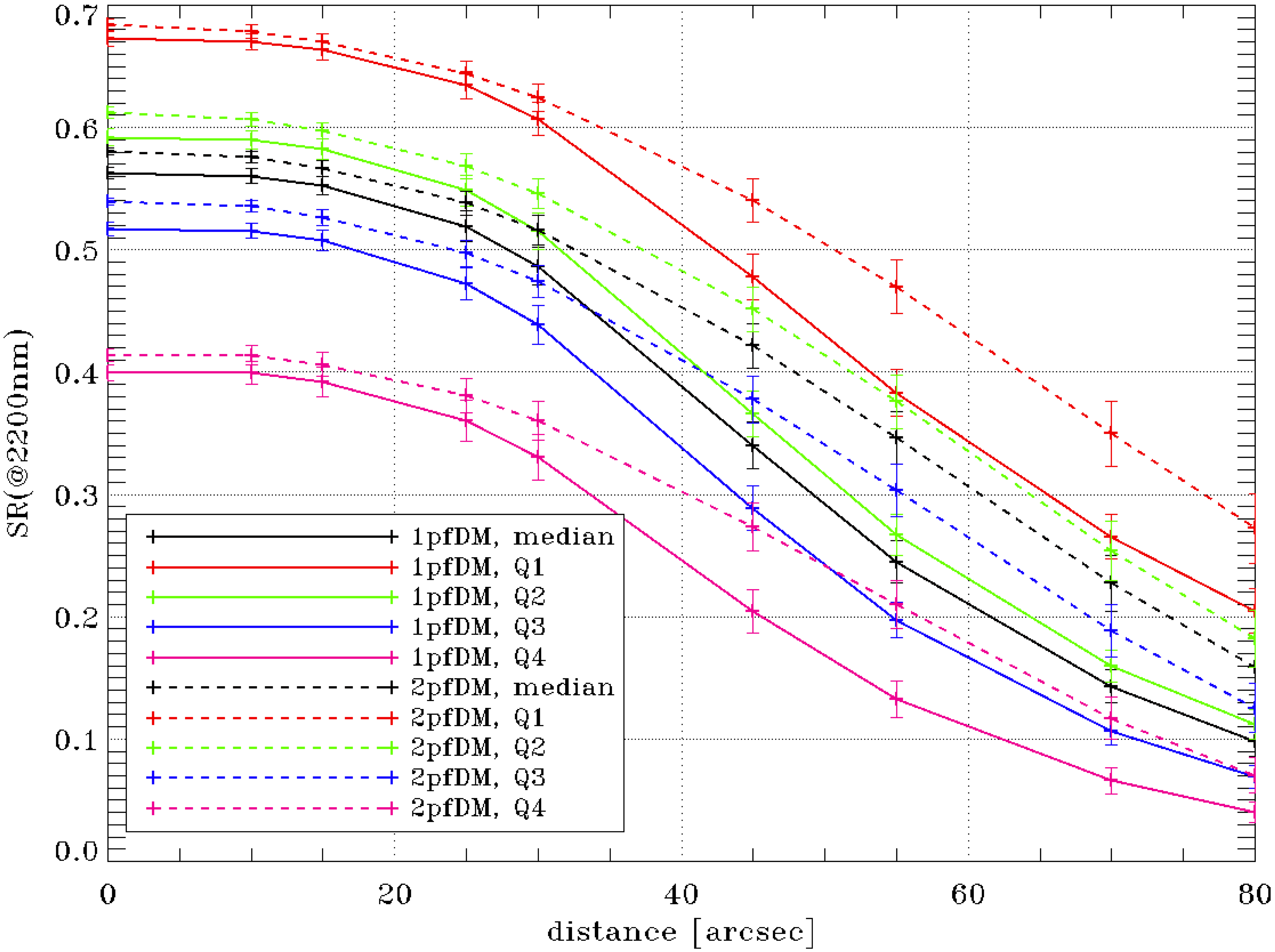}}
\subfigure[On-axis K band PSF with the median profile and a single post focal DM in the first atmospheric realization without extra error terms (see Tab.\ref{Tab:params}).\label{fig:PSF}]
{\includegraphics[width=0.45\columnwidth]{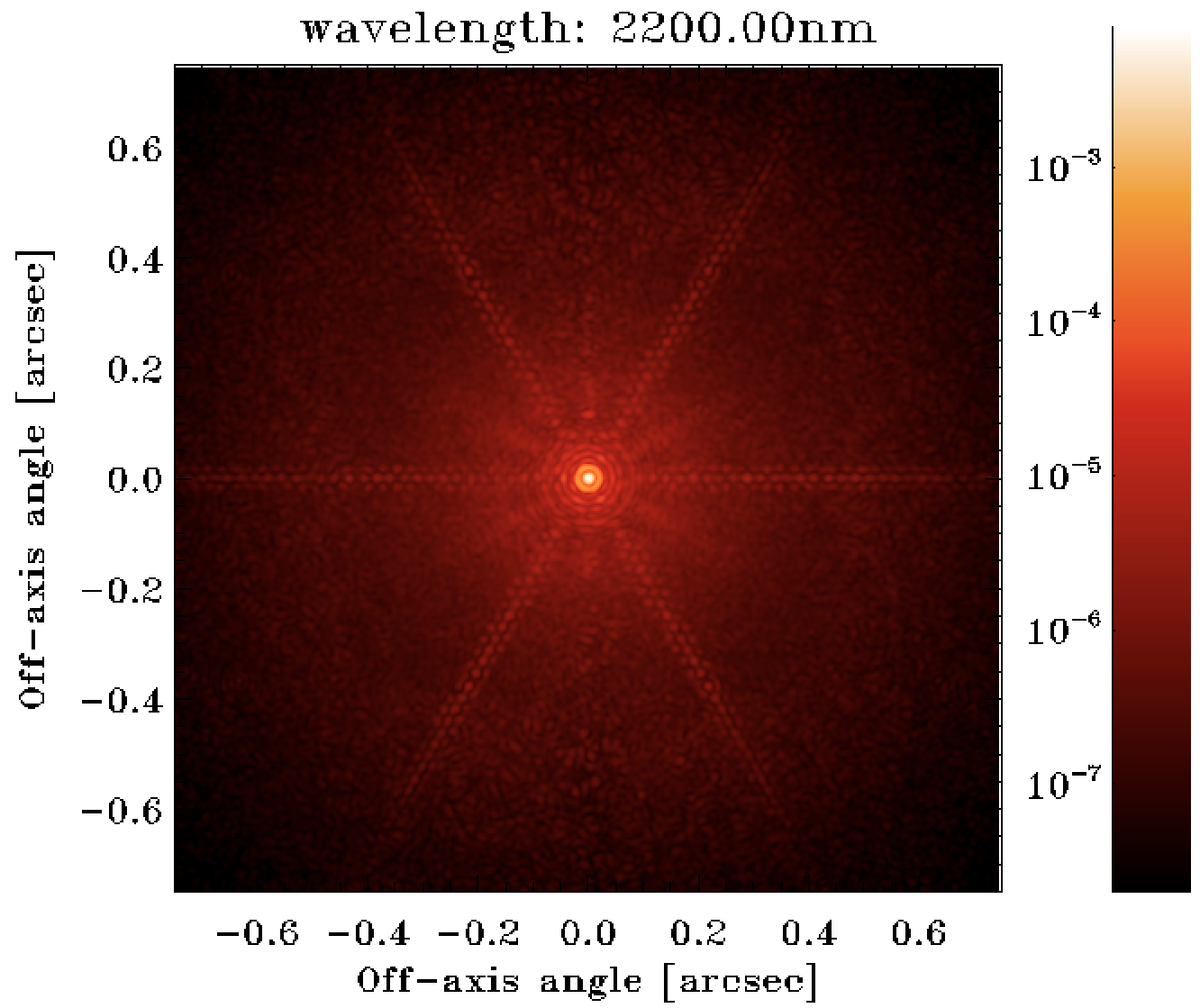}}
\caption{Good NGS asterism results.}\label{fig:goodAst}
\end{figure}

\begin{figure}
    \begin{center}
    \begin{tabular}{c}
        \includegraphics[width=0.65\textwidth]{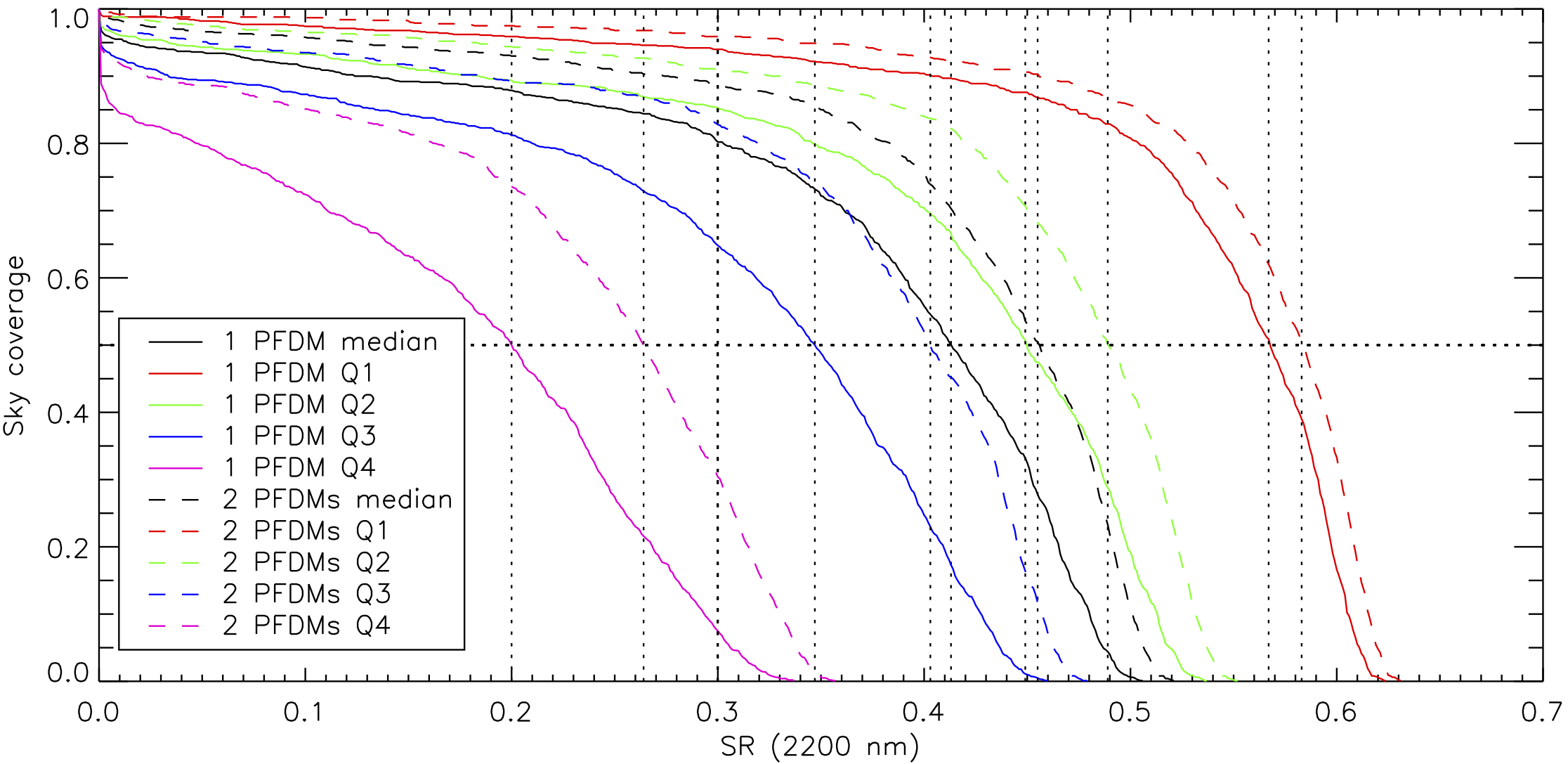}
    \end{tabular}
    \end{center}
	 \caption{\label{fig:sky} Statistical Sky Coverage (average K band SR in the science FoV) for different atmospheric conditions, for a single post focal DM and two post focal DMs.}
\end{figure}

The baseline configuration of the MAORY system that is used in numerical simulation is presented in Tab. \ref{Tab:params}.
The performance in terms of Strehl Ratio (SR) vs Field for a good NGS asterism (bright and close to science field\footnote{NGS off-axis distance is 55 arcsec (minimum distance to avoid vignetting the science field of MICADO\cite{10.1117/12.2311481}, 53$\times$53 arcsec).}) is shown in Fig. \ref{fig:goodAst} together with an example of K band PSF and the performance in terms of statistical sky coverage is presented in Fig. \ref{fig:sky}.
In both cases we plot the average performance computed from 5 simulations of 2 seconds long with different atmospheric and noise realizations.

Note that we use the same reconstruction matrix for all configurations and it is the one optimized for the median atmospheric conditions.
We chose this because we verified that optimizing for the specific condition gives a small (with respect to our error budget) improvement of a few tens of nm.
Instead, projection matrix changes in the case of the 2 post focal DM: in this case, 4 of the 8 optimization directions at 30arcsec are moved to 60arcsec to get a performance improvement in the technical field of view (see Fig. \ref{fig:SR}) that boosts the sky coverage (see Fig. \ref{fig:sky}).

Currently we have a good margin with respect to the requirement of 30\% SR in K band on half of the sky with median atmospheric conditions, and we are able to get close to the goal (50\% SR in K band) with the 2 post-focal DM configuration.
The difference between the two configurations is negligible with the good NGS asterism, but when considering statistics, the 2 post focal DMs offer a few percents more of SR and give a substantial boost to performance for bad seeing cases and high sky coverage ($\sim$80\%).

Data coming from these simulations is used by the science team to build a set of synthetic Point Spread Functions (PSFs) to study a few aspect of the capability of MAORY-MICADO\cite{10.1117/12.2311481} and to compare the single and double post focal DMs case on scientific targets \cite{carmelo2020science}.
\begin{table}[ht]
\caption{Summary of the baseline MAORY parameters (in simulation).}
\label{Tab:params}
\begin{center}
\begin{small}
	\begin{tabular}{|l|c|}
		\hline
		\textbf{Parameter} & \textbf{value}\\
		\hline
		Telescope diameter &  38.5m\\
		Central obstruction & 11.0m\\
		Pupil mask & M1 pupil with spiders\\
		Zenith angle & 30deg\\
		Science FoV (diameter) & 60arcsec\\
		Technical FoV (diameter) & 160arcsec\\
		\hline
		Atmospheric turbulence & 5 profiles with 35 layers (Q1, Q2, Q3, Q4 and median)\\
		\multirow{2}{*}{$r_0$} & 0.234m (Q1), 0.178m (Q2), 0.157m (median),\\
		 & 0.139m (Q3) and 0.097m (Q4)\\
		$L_0$ & 25m\\
		\hline
		Full throughput NGS path (H band) & 0.33\\
		Full throughput LGS path & 0.23\\
		Full throughput ref. path (R+I band) & 0.15\\
		\hline
		Sky background NGS (H band) & 2444$\mathrm{e^-/m^2/arcsec^2/s}$ (no moon)\\
		Sky background reference (R+I band) & 39$\mathrm{e^-/m^2/arcsec^2/s}$ (no moon)\\
		\hline
		Ground DM & conjugated at 600m with $\sim$0.5m pitch (num. inf. func. of M4)\\
		Post focal DM no 1 & conjugated at 17500m with $\sim$1.25m pitch\\
		Post focal DM no 2 (optional) & conjugated at 7000m with $\sim$0.85m pitch\\ 
		\hline
		NGS WFS number & 3\\
		NGS WFS off-axis angle (good ast. case) & 55arcsec\\
		NGS WFS no SA & 2x2\\
		NGS WFS FoV & 1.3arcsec\\
		NGS WFS pixel pitch & 8.6mas\\
		NGS WFS detector RON & 0.5$\mathrm{e^-/pixel/frame}$\\
		NGS WFS detector dark current & 20$\mathrm{e^-/pixel/s}$\\
		NGS WFS thermal background & 3$\mathrm{e^-/pixel/s}$\\
		\hline
		LGS WFS number & 6\\
		LGS WFS off-axis angle & 45arcsec\\
		LGS WFS no SA & 70x70\\
		NGS WFS FoV & 16.8arcsec\\
		LGS WFS pixel pitch & 1.2mas\\
		LGS WFS detector RON & 3.0$\mathrm{e^-/pixel/frame}$\\
		LGS launcher off-axis distance & 21m\\
		\hline
		Sodium profile & ``multi peak''\cite{2014A&A...565A.102P}\\
		\hline
		Control & POLC with split tomography\cite{Busoni2019} and noise priors\cite{Michel-Tallon:2008aa,Oberti2019}\\
		Reconstruction layers altitude &  [0.6, 2.0, 4.5, 7.5, 11.0, 15.0, 18.0, 22.0] km\\
		Reconstruction layers pitch & [0.50, 0.55, 0.60, 0.70, 0.75, 0.80, 0.85, 1.00] m\\
		Optimization directions\cite{Fusco:01} & 1 on-axis, 8 at 30arcsec and 4 at 80 arcsec with low weights\\
		Centroiding algorithm & CoG with smart windowing\cite{Oberti2019} (LGS), weighted CoG (NGS)\\
		Framerate & 500Hz (LGS), 100-500Hz (NGS)\\
		Total delay & 5.5ms\\
		Integrator gains & 0.25 (LGS), 0.1 (NGS)\\
		\hline
		Extra error term & 105nm\cite{Busoni2019}\\
		\hline
		\multicolumn{2}{c}{\scriptsize Note: WFS is Wavefront Sensor, NGS is Natural Guide Star, LGS is Laser Guide Star, DM is Deformable Mirror, SA is}\\
		\multicolumn{2}{c}{\scriptsize Sub-Aperture, FoV is Field of View, RON is Read-Out Noise, POL is Pseudo-Open Loop Control and CoG is Center of Gravity.}\\
	\end{tabular}
\end{small}
\end{center}
\end{table}
%
%

\section{SENSITIVITY ANALYSIS}\label{sec:sens}

A few sets of sensitivity analysis have already been published in Ref.~\citeonline{10.1117/12.2313175,Plantet2019} and \citeonline{Busoni2019}.
These analyses helped the system engineering in the selection of many parameters, like the technical Field of View (FoV)\cite{10.1117/12.2313175,Plantet2019}, the Laser Guide Star (LGS) asterism\cite{Plantet2019}, the number of sub-apertures in the LGS wavefront sensor (WFS)\cite{Busoni2019} and Natural Guide Star (NGS) WFS\cite{Plantet2019} and the post focal Deformable Mirror (DM) pitch\cite{Busoni2019}.
Here we focus on new sensitivity analyses done in the last year.
They involved a few parameters: number of sub-apertures in the LGS WFS, zenith angle, the use of Atmospheric Dispersion Compensators (ADC) in the NGS WFSs, sodium profile and mis-registration of the LGS WFSs and of the DMs.

Unless specified otherwise, we consider a single post focal DM and a bright NGS asterism at 55 arcsec on an equilateral triangle.

\subsection{Number of sub-apertures with super-resolution}\label{sec:noSA}

We started to explore the possibility to introduce calibrated shifts and rotations in the LGS WFSs to get the ability to super-resolve the incoming wavefront.
Super-resolution is a well known feature in imaging and it has already been tested on a AO system (see Ref. \citeonline{Woillez2019} and Oberti et al., forthcoming).
We are already aware of the super-resolution at the conjugation altitude of M4 (about 600m) given by the different lines of sight of the 6 LGSs\cite{Busoni2019}, but now we want to improve the resolution close to the pupil where most of the turbulence is concentrated.
This can be done by deliberately shifting and/or rotating the WFSs with respect to the pupil.
The sensing pitch can be increased by choosing the correct shifts and rotations scheme, but resolution cannot be increased on the whole range of altitudes because of the different lines of sight of the LGSs and the partial coverage of the metapupil at the highest altitudes.

\begin{figure}[h!]
\centering
\subfigure[K band SR as a function of the off-axis angle for different LGS WFSs configurations.\label{fig:superError}]
{\includegraphics[width=0.5\columnwidth]{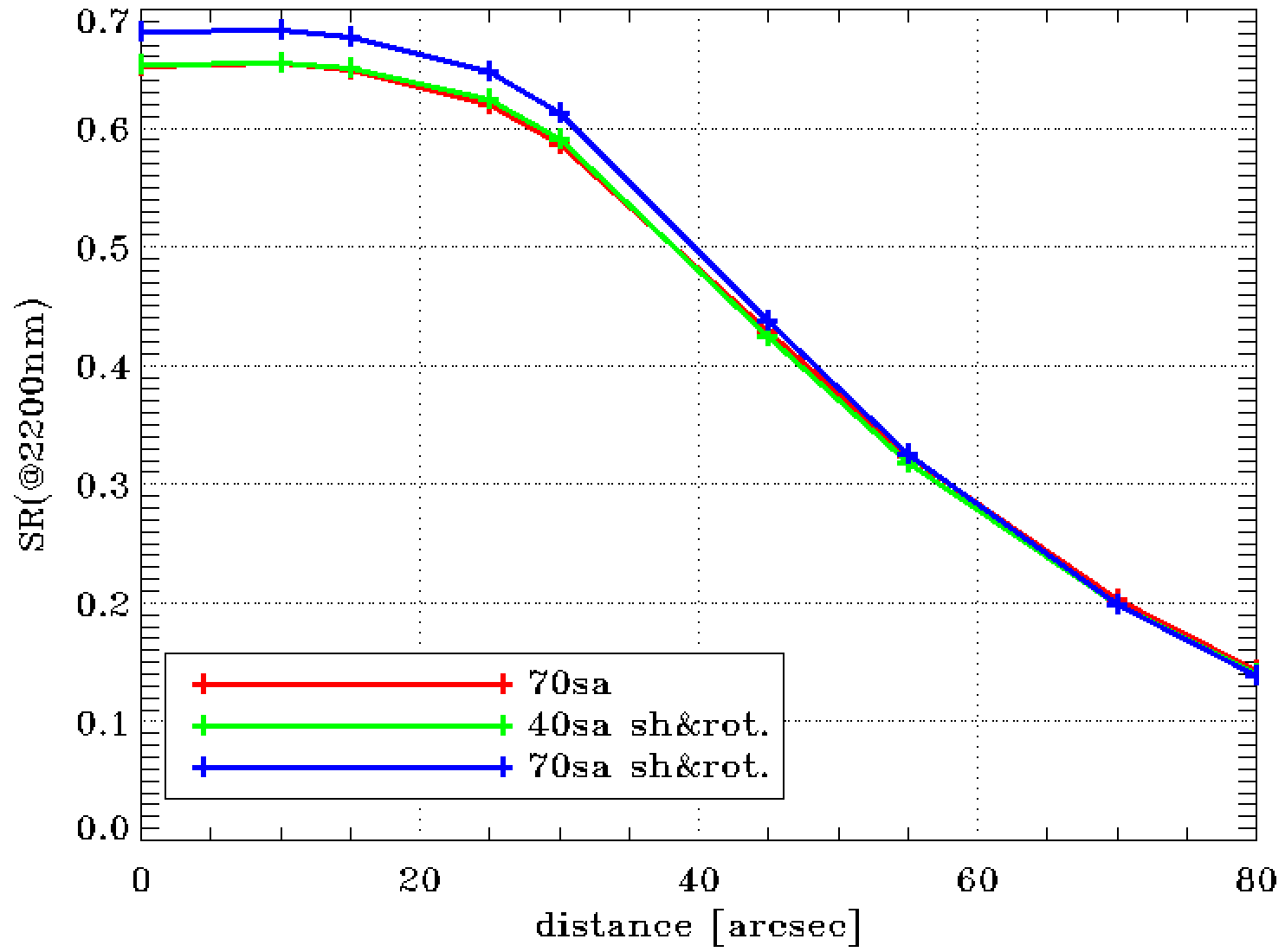}}
\subfigure[Shifts and rotations scheme.\label{fig:superShAndRot}]
{\includegraphics[width=0.3\columnwidth]{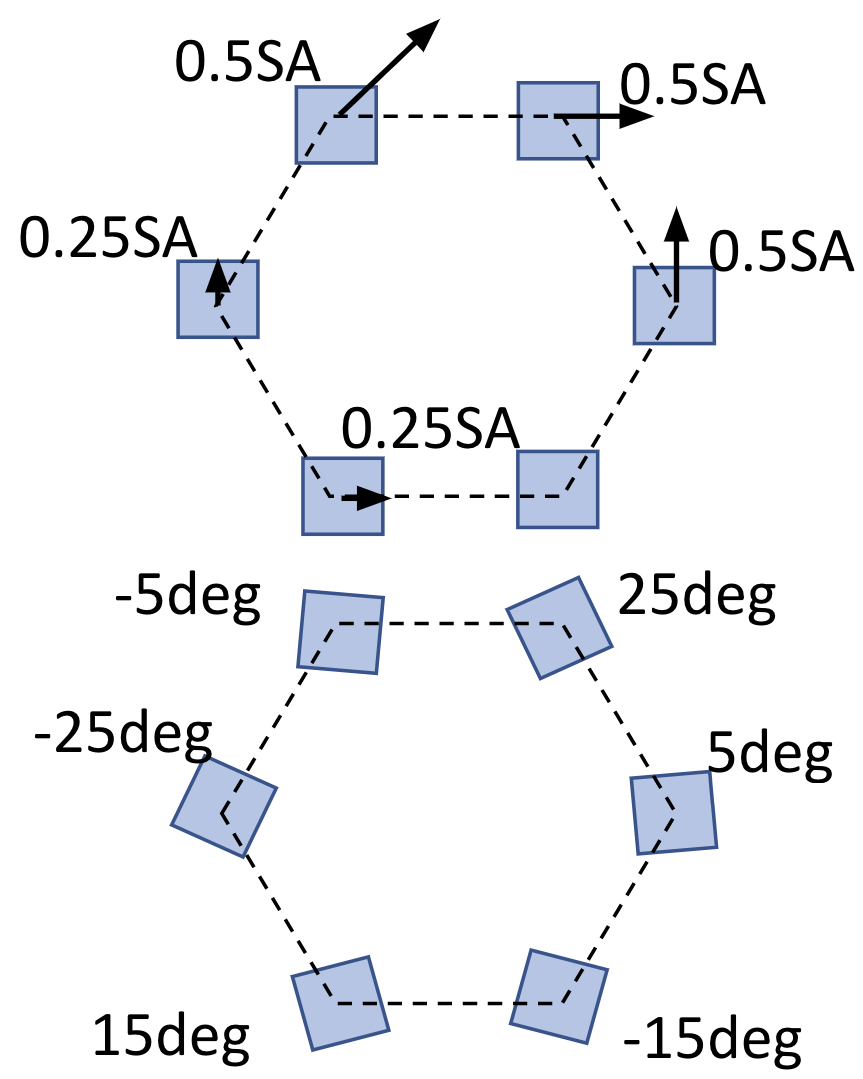}}
\caption{Super-resolution and number of sub-apertures.}\label{fig:super}
\end{figure}

Here we present our preliminary results done on a couple of LGS WFSs configuration with the baseline number of sub-apertures, 70, and on a greatly reduced one with just 40 sub-apertures.
Results are shown in Fig. \ref{fig:superError} and the WFS shifts and rotations scheme in \ref{fig:superShAndRot}.
Super-resolution is really effective in improving residual error of the system, giving an equivalent performance on a super-resolved 40$\times$40 sub-apertures case and on a ``classic'' 70$\times$70 sub-apertures.
Moreover, it improves the performance of a configuration with a sub-aperture dimension close to the DM pitch as the 70$\times$70 sub-apertures. 

Note that the WFS shifts and rotations scheme in \ref{fig:superShAndRot} is not optimized, it is a first attempt.
We are currently working on finding a way to optimize this scheme, but it is not trivial because of the geometry of the system, with different lines of sight and a conical propagation.
In fact, the optimization can be done easily on a single altitude of the sensed volume, but this does not guarantee the same super-resolution on different altitudes.

%

\subsection{Zenith angle and atmospheric dispersion}\label{sec:zenith}

\begin{figure}[h!]
\centering
\subfigure[1 post focal DM case.\label{fig:1dmsky45}]
{\includegraphics[width=0.49\columnwidth]{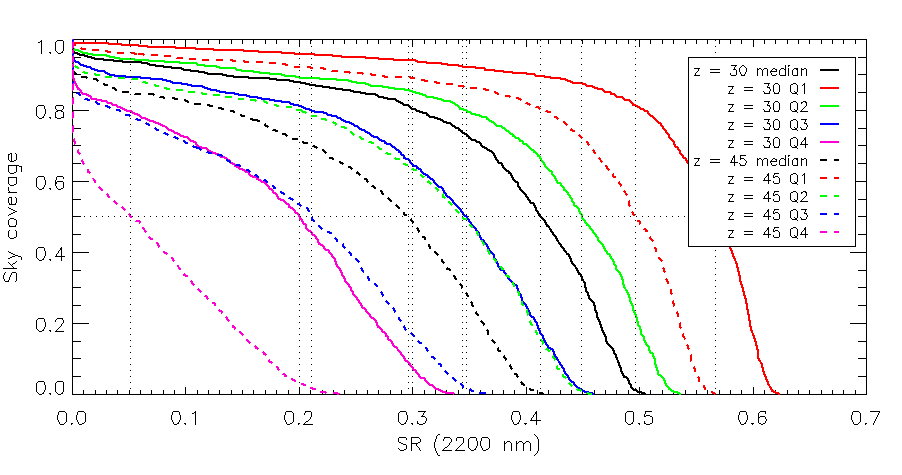}}
\subfigure[2 post focal DMs case.\label{fig:2Dmsky45}]
{\includegraphics[width=0.49\columnwidth]{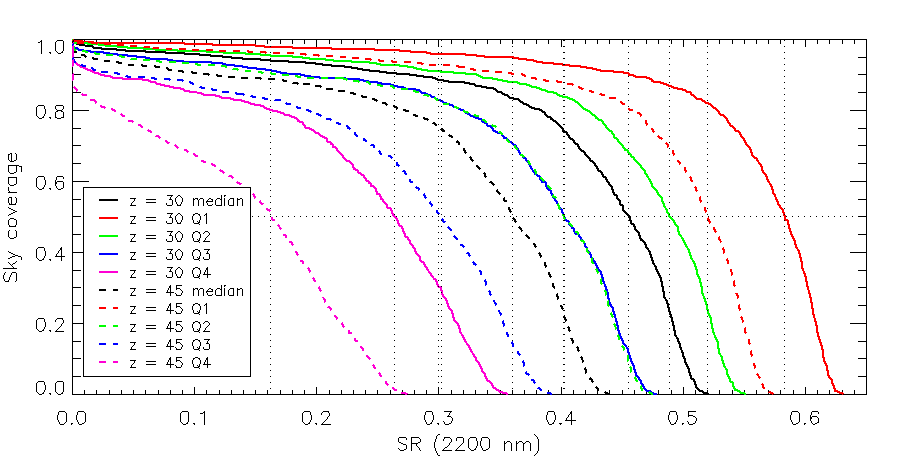}}
\caption{Sky coverage with different elevation angles.}\label{fig:sky45}
\end{figure}

We studied the impact on sky coverage when the zenith angle is 45 degrees.
The results are shown in Fig. \ref{fig:sky45}.
In this case we had to compute the sky coverage because the impact of zenith angle is not limited to the LGS correction, but also to the NGS one.
We see that the impact on performance is important and larger than the one expected for the effect of the airmass on the seeing.
Indeed, the airmass also changes the anisoplanatic angle $\theta_0$, which is strongly correlated to the performance of a MCAO system.

Note that in this case reconstruction matrix was recomputed for the different zenith angle because LGS footprints on the reconstruction layers change in size and Cn$^{2}$ altitudes scale with the airmass.
Instead projection matrices are the same because altitude of reconstruction layers (their altitude, see Tab. \ref{Tab:params}, was chosen to be compatible with both 30 and 45deg of zenith angle) and DMs do not change.

\begin{figure}
    \begin{center}
    \begin{tabular}{c}
        \includegraphics[width=0.75\textwidth]{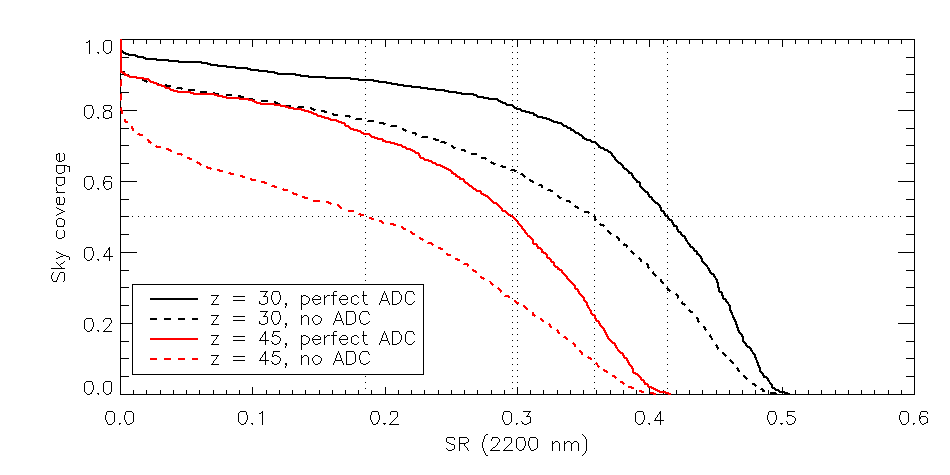}
    \end{tabular}
    \end{center}
	 \caption{\label{fig:ADC} Sky coverage for median atmospheric condition and a single post focal DM with different elevation angles and with (solid lines) and without atmospheric dispersion (dashed lines) on the NGS WFSs.}
\end{figure}

Then we replicate this study in median atmospheric conditions, considering the atmospheric dispersion on the NGS WFS (2$\times$2, H band).
This is done using a polychromatic model of the Shack-Hartmann WFS (standard version is monochromatic) with 5 wavelengths, 1510, 1570, 1.630, 1690 and 1750nm, and different tilts.
These results, shown in Fig. \ref{fig:ADC}, prove that an ADC will significantly improve the performance. 


\subsection{Sodium profile}

\begin{figure}[h!]
\centering
\subfigure[K band SR as a function of the off-axis angle for the ``multi-peak'' and ``very wide'' sodium profiles.\label{fig:sodiumError}]
{\includegraphics[width=0.48\columnwidth]{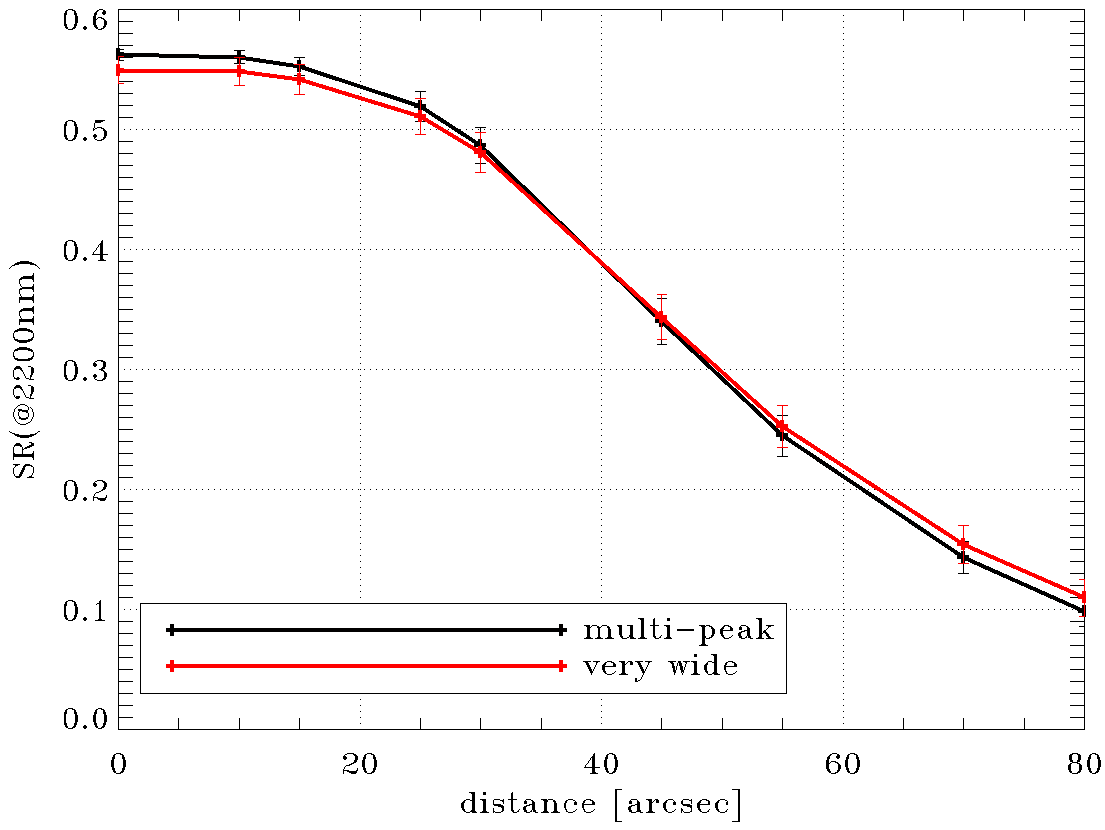}}
\subfigure[On-axis RMS of turbulence and residual projected on a modal base (made by 5 Zernikes and 4495 Karhunen-Lo{\'e}ve modes\cite{Wang:78}) with and without truth sensing in the case of a ``very wide'' sodium profile.\label{fig:truth}]
{\includegraphics[width=0.49\columnwidth]{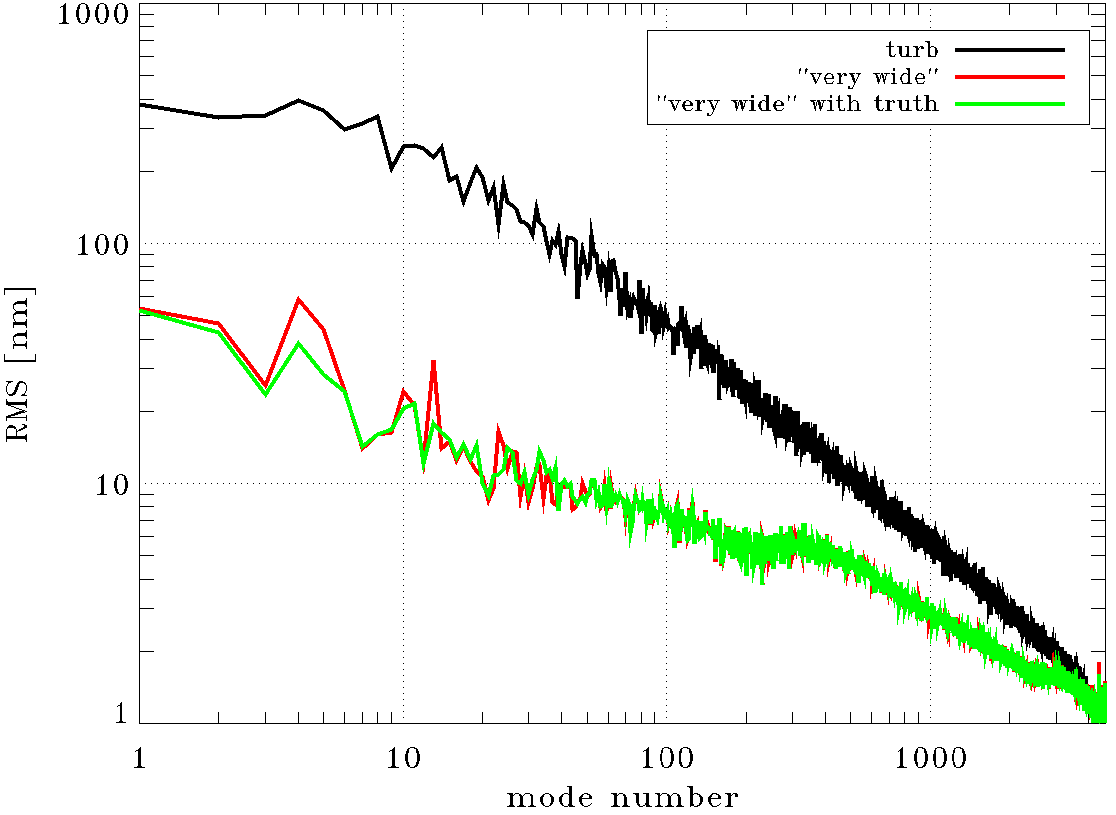}}
\caption{Sodium elongation and performance.}\label{fig:sodimElongation}
\end{figure}

We tested the system with three different sodium profiles from Ref.~\citeonline{2014A&A...565A.102P}: ``multi-peak'', ``top hat with peak'', and 
``very wide''.
We limited the study to a good NGS asterism because we are interested in the LGS correction.
Thanks to our control approach, the sensitivity to sodium profile variations is relatively low: the first two profiles give an almost identical performance and only the ``very wide'' sodium profile gives a lower performance by $\sim$1\% K band SR in the science FoV as shown in Fig. \ref{fig:sodiumError}.

Note that priors\cite{Michel-Tallon:2008aa,Oberti2019} were selected for each specific sodium profile so reconstruction matrix changes, but using the one for the ``very wide'' sodium profile for all the cases gives a small impact of about 1\% of K band SR on science FoV for the best sodium profiles.
Moreover the windows used for the center of gravity computation are optimized on the specific sodium profile, but this can be done live on the average flux detected on the sub-apertures.

Then, we evaluate the capability of MAORY truth sensing to recover the bias induced by the truncation error.
Truth sensing is tomographic, slow (integration time 1s and bandwidth $\sim$0.01Hz) and it is given by 3 Shack-Hartmann WFSs (10$\times$10 sub-apertures with 24$\times$24 pixel of 0.167arcsec/pixel) working with the visible light (600-1000nm) of the NGSs\cite{10.1117/12.2313266,Busoni2019}. It is clear from the results shown in Fig. \ref{fig:truth} that truth sensing is able to reduce residual on those modes where truncation induced errors are stronger (mainly modes 4 and 5 that are the two astigmatisms). 
%
%
\subsection{Mis-registrations}\label{sec:misreg}

Using the same feature developed for super-resultion, we evaluated the sensitivity to mis-registration between the LGS WFSs and the pupil.
We add a non-calibrated shift or rotation on one of the 6 WFSs by various amounts and we compared the performance with the nominal case.
We replicated this for the super-resolved configurations presented in Sec. \ref{sec:noSA} (in this case we have a portion of the shifts and rotations that is calibrated in the reconstruction matrix, the one shown in Fig. \ref{fig:superShAndRot}, and a part that is not) and we saw that the sensitivity with these configurations is reduced, in particular for the one with less sub-apertures (as expected).

Then we consider a mis-regitration between the DMs and the pupil (hence WFSs).
The results show that we can tolerate a shift of about 0.2 times the DM pitch before getting an unstable performance.
\begin{figure}[h!]
\centering
\subfigure[Error as a function of shifts of a single WFS along the x direction (the three WFSs configurations of Sec.\ref{sec:noSA} are used).\label{fig:errShift}]
{\includegraphics[width=0.4\columnwidth]{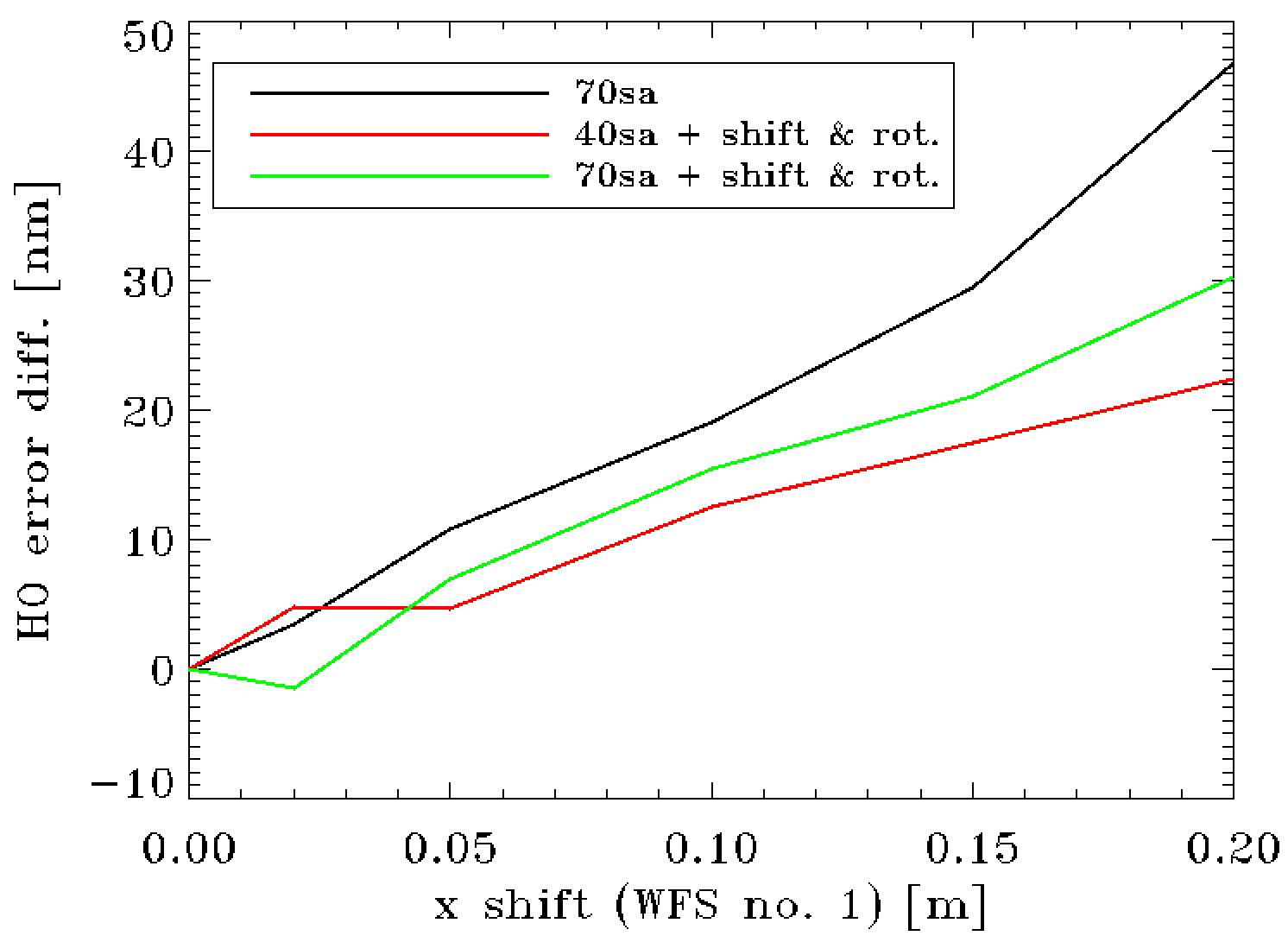}}
\subfigure[Error as a function of rotations of a single WFS (the three WFSs configurations of Sec.\ref{sec:noSA} are used).\label{fig:errRot}]
{\includegraphics[width=0.4\columnwidth]{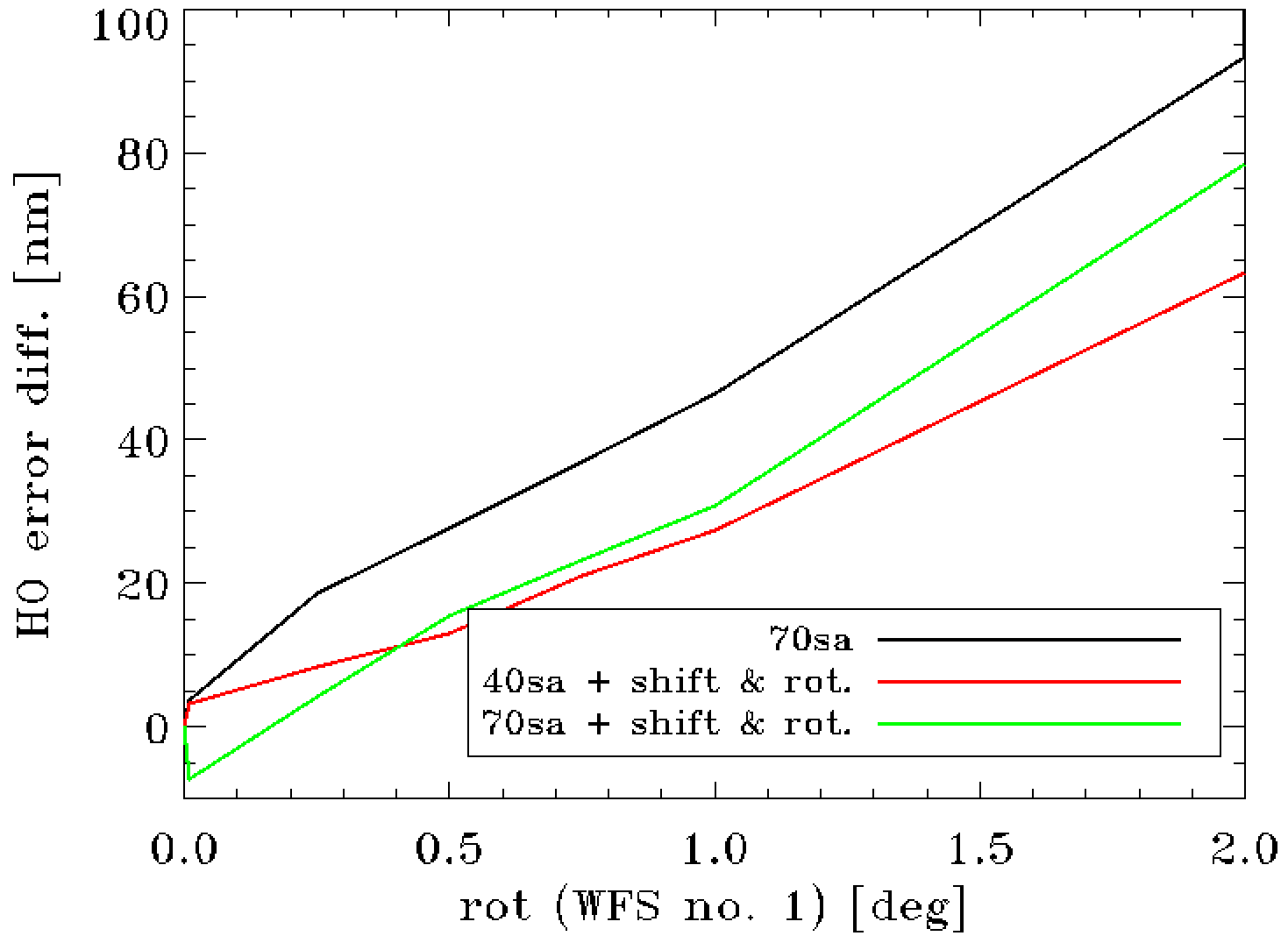}}
\subfigure[Error as a function of shifts of M4 along the x direction.\label{fig:errShiftM4}]
{\includegraphics[width=0.4\columnwidth]{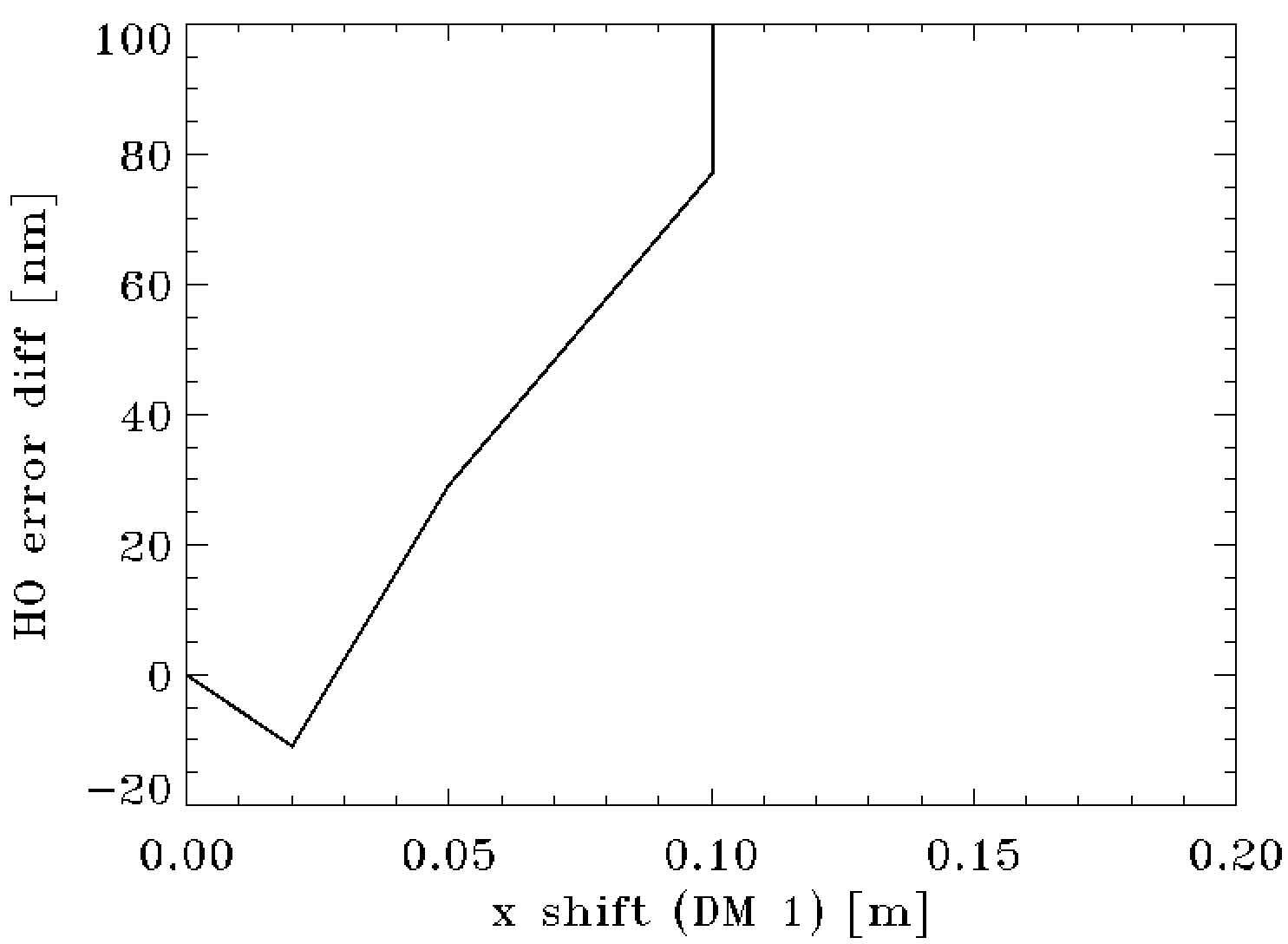}}
\subfigure[Error as a function of shifts of the post focal DM along the x direction.\label{fig:errShiftPost}]
{\includegraphics[width=0.4\columnwidth]{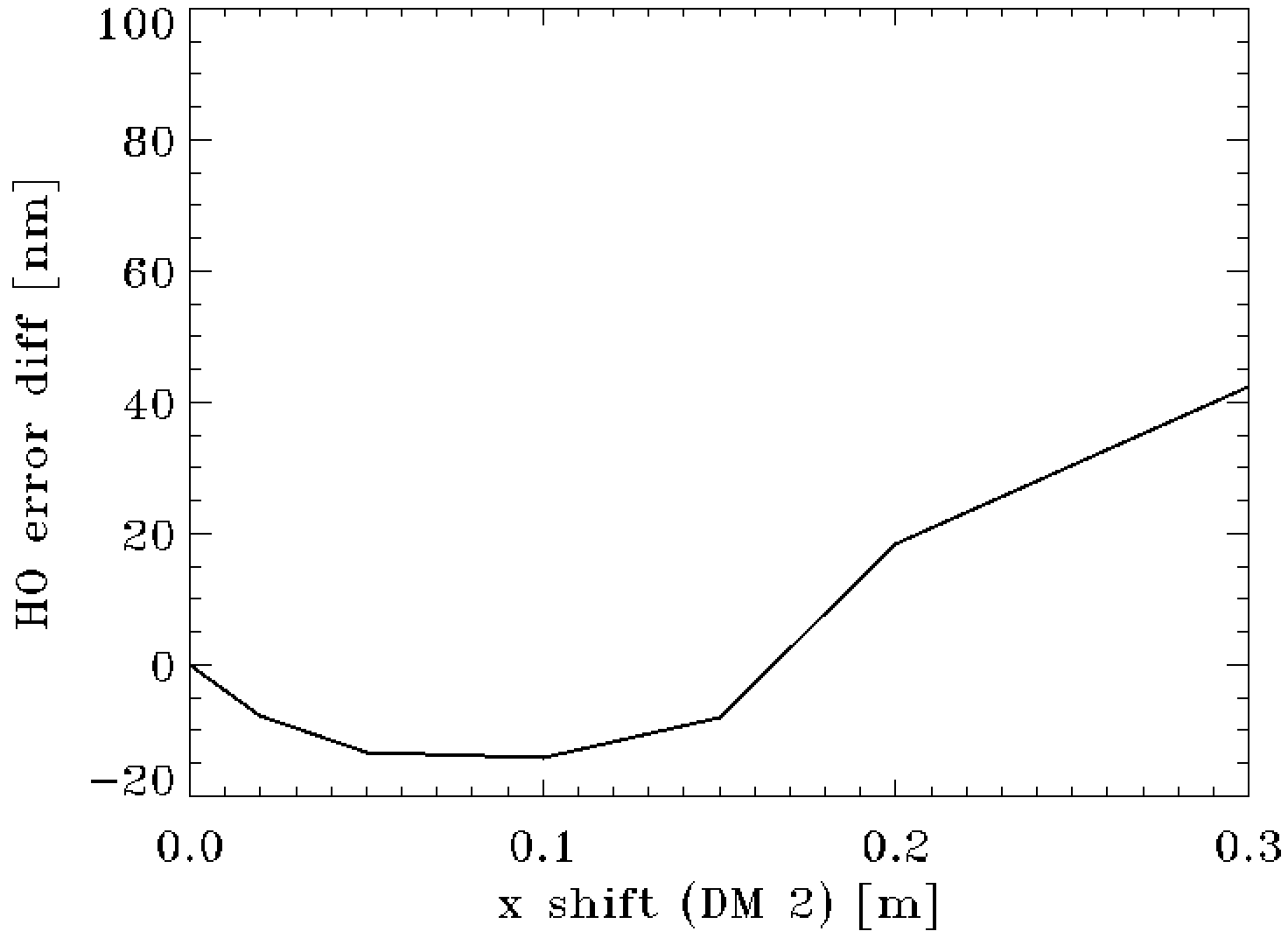}}
\caption{Sensitivity analysis for mis-registrations.}\label{fig:misReg}
\end{figure}

\section{Conclusion}

We showed the estimated performance of the MAORY system and its sensitivity to many parameters.
This work has been coordinated with the activity of engineering and design of MAORY and helped in the configuration of the current baseline.
This baseline (depicted in Tab. \ref{Tab:params}) is able to reach the requirements with a good margin with both a single and two post-focal DMs.
The two post focal DM gives better performance in all conditions, as expected, and in particular in the most demanding ones.

Now we are focusing on the incoming preliminary design review programmed for the beginning of 2021: we are currently completing all analysis and working on a way to compare the impact on science operations of the two post focal DM configurations to get a more pragmatic information.

\appendix
\section{PASSATA for MAORY}\label{sec:appendix}
In the MAORY framework, PASSATA has been revised to reduce memory requirements and to improve its speed in addition to adding all the new required features like the ones described in Sec. \ref{sec:zenith} and \ref{sec:misreg}.
Hence, it is able to run more complex simulations without requiring a significant hardware update of the workstation used for Single Conjugate AO systems.

This has mostly involved the way simulations are carried out, but also the DLM extension based on CUDA functions.

On the simulation side, we chose to divide a simulation in multiple steps.
The idea is to separate the wavefront correction and the performance estimation:
\begin{itemize}
    \item In a first step we run the closed loop between input disturbances (atmospheric turbulence, structure vibrations, \dots), WFS and DMs but we avoid computing the PSF and other performance metrics so that both memory requirements and time spent decrease.
    \item In a second step we load the same input disturbance realization (identified by a seed value) and the same correction (defined by combinations of DM parameters and DM command history saved in the previous step) from the disk, so that we are able to replicate the same residual wavefront and compute PSFs and/or other performance metrics. Since many elements of the simulations are not used in this step, like WFSs and reconstruction, memory requirements and time spent are kept low. Moreover, this step can be replicated many times computing PSFs at different positions or wavelengths.
\end{itemize} 

Moreover this multiple step approach is useful in MCAO simulations with a split control between NGS and LGS, as it is now for MAORY, because different NGS asterism simulations can be run using a single initial step where only LGS correction is computed as it described in Ref. \citeonline{Busoni2019}.

Then, on the CUDA functions side, we mainly focused on Shack-Hartmann WFS implementation to minimize its memory consumption sharing part of the memory between sensor with the same parameters. 

\acknowledgments

The  MAORY  instrument  is  developed  in  the  framework  of  the  Agreement  No.  65221/ESO/15/67001/JSC  between  the  European  Organisation 
for  Astronomical  Research  in  the  Southern  Hemisphere  (ESO)  and  the  Istituto  Nazionale  diAstrofisica  (INAF)  on  behalf  of  the  Consortium  
consisting  of  INAF,  Institut  National  des  Sciences  de  I'Univers du Centre National de la Recherche Scientifique (INSU/CNRS) acting on behalf 
of the Institut de Planetologie et d'Astrophysique de Grenoble (IPAG) and of  School of Physics at the National University of Ireland Galway (NUIG). 

\bibliography{biblio}
\bibliographystyle{spiebib}

\end{document}